\documentclass[sigplan]{acmart}
\setcopyright{none}
\renewcommand\footnotetextcopyrightpermission[1]{}

\AtBeginDocument{%
  \providecommand\BibTeX{{%
    \normalfont B\kern-0.5em{\scshape i\kern-0.25em b}\kern-0.8em\TeX}}}

\usepackage[ruled,vlined,linesnumbered]{algorithm2e}
\usepackage{booktabs}
\usepackage{graphicx}

\begin{document}

\title{Looney Tunes: Exposing the Lack of DRM Protection in Indian Music Streaming Services}

\author{Ahaan Dabholkar}
\email{ahaand@iitbhilai.ac.in}
\affiliation{%
\institution{de.ci.phe.red Lab, Indian Institute Of Technology Bhilai}
}

\author{Sourya Kakarla}
\email{sourya4@iitkgp.ac.in}
\affiliation{%
\institution{Indian Institute Of Technology Kharagpur}
}

\author{Dhiman Saha}
\email{dhiman@iitbhilai.ac.in}
\affiliation{%
\institution{de.ci.phe.red Lab, Indian Institute Of Technology Bhilai}
}

\begin{abstract}
    Numerous studies have shown that streaming is now the most preferred way of consuming multimedia content and this is evidenced by the proliferation in the number of streaming service providers as well as the exponential growth in their subscriber base. Riding on the advancements in low cost electronics, high speed communication and extremely cheap data, Over-The-Top (OTT) music streaming is now the norm in the music industry and is worth millions of dollars. This is especially true in India where major players offer the so called freemium models which have active monthly user bases running in to the millions. These services namely, Gaana\cite{gaana}, Airtel Wynk\cite{wynk} and JioSaavn\cite{jiosaavn} attract a significantly bigger audience than their 100\% subscription based peers like Amazon Prime Music, Apple Music etc.\cite{ecotimesreport} Given their ubiquity and market dominance, it is pertinent to do a systematic analysis of these platforms so as to ascertain their potential as hotbeds of piracy. This work investigates the resilience of the content protection systems of the four biggest music streaming services (by subscriber base) from India, namely Airtel Wynk, Ganna, JioSaavn and Hungama. By considering the Digital Rights Management (DRM) system employed by Spotify as a benchmark, we analyse the security of these platforms by attempting to steal the streamed content efficiently.
    Finally, we present a holistic overview of the flaws in their security mechanisms and discuss possible mitigation strategies. To the best of our knowledge, this work constitutes the first attempt to analyze security of OTT music services from India. Our results further confirm the time tested belief that security through obscurity is not a long term solution and leaves such platforms open to piracy and a subsequent loss of revenue for all the stakeholders.
\end{abstract}


\maketitle
\pagestyle{plain}

\section{Introduction}
\label{introduction}

OTT is an acronym for ``over-the-top'' and refers to the distribution of multimedia (audio, video) content over a public network. Recent trends have shown a mass adoption of smart mobile devices in the consumer market. This coupled with a higher penetration of high-speed, cheap Internet and the emergence of advanced technologies, such as 5G, 4G, developed online payment infrastructure and continual demand for content within the entertainment domain, projects the global OTT service market to grow from \$81.60 billion in 2019 to \$156.9 billion by 2024 exhibiting a CAGR (Compound Annual Growth Rate) of 14\% ~\cite{mnm20}. The Asia Pacific region is set to record the highest growth rate during the forecast period. According to a joint report published by the Indian Music Industry (IMI) and Deloitte India~\cite{ref_deloitte1}, the audio-video OTT market in India is valued at around US\$ 280 million with nearly 150 million monthly active users accessing soundtracks across various platforms.  

\begin{table}[h]
	\centering
	\resizebox{\linewidth}{!}{
		\renewcommand{\arraystretch}{1.5}
		\begin{tabular}{cccc}
			\hline
			\textbf{Service Name} & \textbf{Business Model}                                         & \textbf{Origin} & \textbf{Reference}  \\ \hline
			Airtel Wynk           & \begin{tabular}[c]{@{}c@{}}Bundle, \\ Ad Supported\end{tabular} & Domestic        & \cite{wynk}  \\ \hline
			Apple Music           & Paid                                                            & International   & \cite{apple-music}  \\ \hline
			Amazon Music          & Paid                                                            & International   & \cite{amazon-music} \\ \hline
			Gaana                 & Ad Supported                                                    & Domestic        & \cite{gaana}        \\ \hline
			Hungama               & Ad Supported                                                    & Domestic        & \cite{hungama}      \\ \hline
			JioSaavn              & \begin{tabular}[c]{@{}c@{}}Bundle,\\ Paid\end{tabular}          & Domestic        & \cite{jiosaavn}     \\ \hline
			Spotify               & Ad Supported                                                    & International   & \cite{spotify_ref}      \\ \hline
			Youtube Music         & Subscription                                                    & International   & \cite{youtube-music} \\ \hline
		\end{tabular}
	}
	\caption{OTT music services currently operating in India}
	\label{tab:my-table}
\end{table}

Revenue from digital means contributes nearly 78\%to the overall recorded music industry revenue in India and 54\%~\cite{ifpi18}, globally. A survey of India's audio streaming market reveals that it is primarily divided among domestic players Wynk, Gaana, JioSaavn, Hungama and global players Spotify, Amazon Music, Apple Music and more recently YouTube Music (Table~\ref{tab:my-table}). As per a consumer insights survey conducted by the IFPI in 2018~\cite{ifpi_mcir_18}, an average internet user in India spends 21.5 hours every week listening to music, higher than the global average of 17.8 hours. It is interesting to note that despite the popularity, contemporary literature lacks security analysis of \emph{any} of the domestic OTT platforms and forms the primary motivation of this work. 

\begin{figure}
	\centering
	\includegraphics[width=\linewidth]{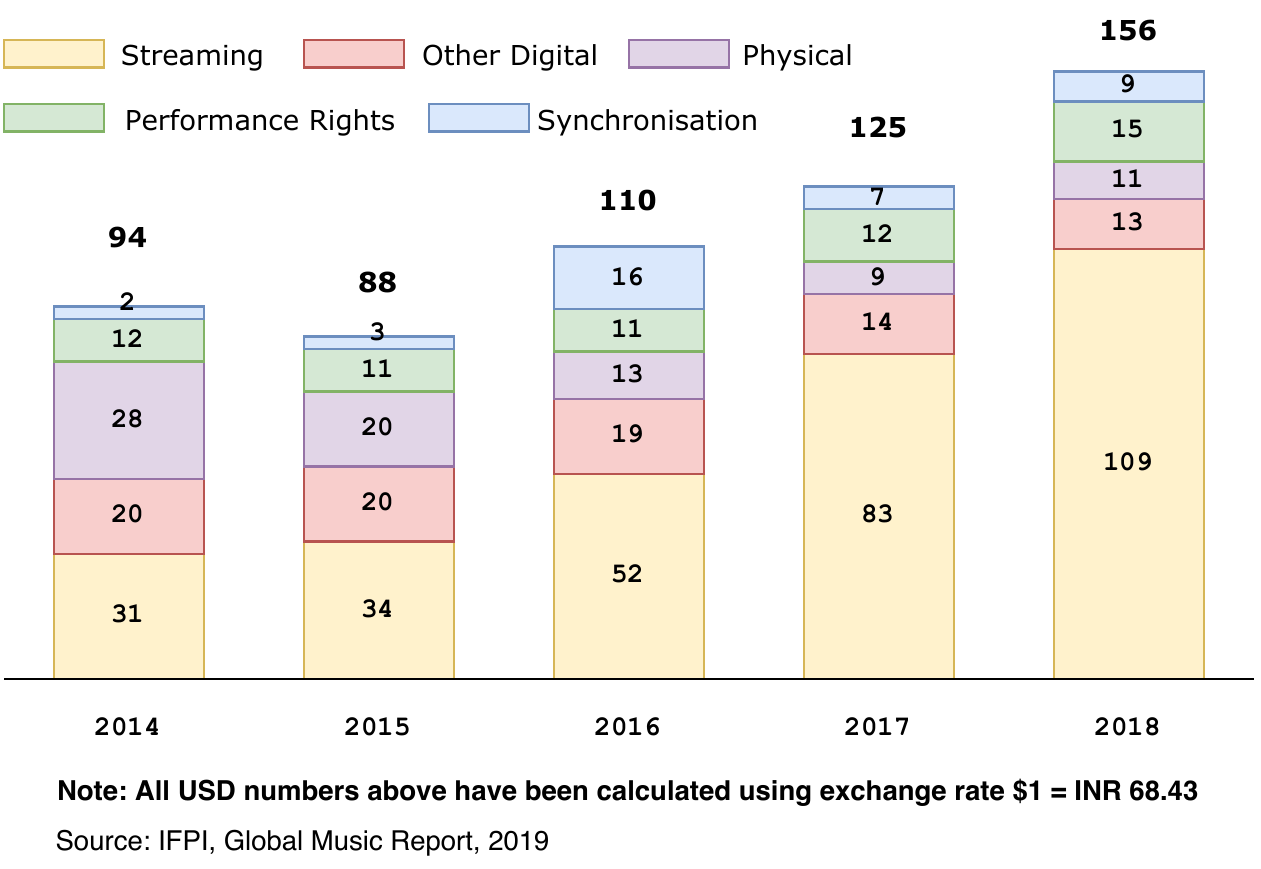}
	\caption{The dominance of streaming as the main source of revenue in the Indian music industry~\cite{ifpi_gmr_19}}
	\label{fig:music-streaming-services-in-india}
\end{figure}

This easy and free access to content was thought to have solved several issues regarding unsanctioned sharing of media~\cite{spotify} as it provided Music-as-A-Service which was more lucrative to the consumer than content ownership~\cite{master_thesis}. However, with the consequential emergence of ``stream-ripping'', piracy has increasingly kept pace. The gravity of the situation reflects in the numbers where estimates point to almost US\$ 250 million lost each year in India alone while the estimated number of stream-rippers in the US have grown to an alarming 17 million~\cite{ref_streamrip}. The surging popularity of such platforms has also not been missed by the shadier sections of our society with more sinister agendas~\cite{ref_blackhat}. Couple this with the 40\% - 60\% of revenue that is lost to pirates, there is hence a dire need to take a critical look at the security of such content delivering platforms. A recent paper on bypassing DRM protection in online video streaming~\cite{DBLP:conf/uss/WangSKV13} is one of the many research efforts highlighting the need to have a deeper understanding on how OTT services should be deployed in practice.

In this work we systematically analyze the four leading OTT music service providers in India namely Wynk, Gaana, JioSaavn and Hungama comparing them to the best practices in the industry. To our great surprise, our research reveals that none of these platforms adopt \textit{any} state-of-the-art DRM protection. Contrary to this they actually attempt a very rudimentary form of code obfuscation. 
As a result, we were able to not only reverse engineer their protocols but also devise mechanisms leading to automated, unsupervised and uninterrupted download of music from their servers. We develop detailed Proof-of-Concepts for the same and illustrate case-studies on each of the platforms. To put things in context, we also investigate the Spotify web-application and find it adopting very standard DRM protection making it a benchmark in the comparative study that we furnish later. Finally, we discuss possible mitigation strategies to salvage the situation.
As a part of responsible disclosure, this work was attempted to be communicated to the concerned parties. With the exception of Wynk, response from others is awaited.

\subsection*{Our Contributions}
Our contributions can be summed up as follows -
\begin{itemize}
    \item We present a security analysis of the content protection systems in place for four of the biggest music streaming services (by subscriber base) in India.
    \item We highlight the lax security protocols in place in \textbf{all} these services by attempting to steal content in an undetectable way and provide proof of concepts to automatically acquire content by reverse engineering their content delivery protocols.
    \item We present a comparative study of these apps with the current state-of-the-art DRM systems.
    \item We present a discussion on the design choices employed by these services and make recommendations to enhance their security.
\end{itemize}
\subsection*{Organisation Of The Paper}
The following sections contain the conclusions and results of our experiments while reverse engineering said services. We first provide a primer on Adaptive Streaming in Section[\ref{adaptive-streaming}] which is used by most of the OTT streaming services and which would help us elucidate the protocols involved clearly. We follow this up in Section[\ref{drm}] with a brief note on present day DRM systems. Section[\ref{spotify}] is dedicated to describing the Widevine DRM used by Spotify to protect it's content, to establish a benchmark for comparing the other services. This leads us into the results of the reverse engineering in Section[\ref{case-study}] where we give reconstructions of the protocols used. Section[\ref{discussions}] contains discussions on the flaws in current DRM systems and the design choices made by these services followed by our conclusions in Section[\ref{conclusions}].
\subsection{Responsible Disclosure}
\label{disclosure}

All the services mentioned here were contacted prior to submission of this manuscript with reports on the vulnerabilities in their protocols and with offers to collaborate on the fix. It should be noted that none of these services have vulnerability disclosure programs and hence finding a suitable point of contact was tough. When informed of the break, Airtel Wynk was all for the idea of a collaborative fix but ended up deploying a haphazard patch without consultation and proper notice which was broken eventually using the same techniques.
\section{Background}
This section is provided as a primer for familiarising the reader with certain technologies that are heavily referred to in this work.
\subsection{Adaptive Streaming}
\label{adaptive-streaming}

Classical streaming protocols used a technique called \textit{progressive streaming} to deliver content. In this technique, a single file sitting on the vendors's server was delivered to the client requesting it. Though this method was simple, it had some obvious inefficiencies which are demonstrated using a toy example below-
\begin{enumerate}
    \item Consider two clients with two different displays, one having a 720p display and the other having a 4K one. With the progressive streaming protocol, both clients would be delivered the \textit{same} content despite the differences in their hardware capabilities. If the content streamed was in 4K, it would not pose a problem for the second client, however for the first client it would imply that he receive 4K media which would eventually be downscaled to 720p (or not run at all, depending on the decoding hardware)
    \item A problem would also arise if one of the clients had severley limited network bandwidth. This client would be unable to consume content meaningfully owing to it's unnecessarily large size.
\end{enumerate}

The idea of \textit{adaptive streaming} aims at solving both these issues in real time. The first problem is solved by having encodings at multiple bitrates of the same media on the content-delivery servers while the second problem is solved by providing the client with the ability to switch between encodings \textit{mid-stream} depending on it's resources. This \textit{adaptiveness} is facilitated by dividing the source content into chunks and indexing it. Hence if network degradation is detected by the client, the next chunk can be retrieved from a lower bitrate, thus maintaining the flow of content.

We will now describe the HTTP Live Streaming (HLS) Protocol, an adaptive streaming protocol which is predominantly used by most streaming services including, in our case, by Wynk Music to serve content efficiently. An understanding of its architecture is necessary to grasp the underlying protocol. The HLS Protocol was developed by Apple Inc and released to the public in 2009. According to survey reports from 2019, HLS remains the most adopted protocol with more than 45\% of broadcasters using it to provide streaming services to their clients.\cite{hls-pop}. 
\subsection{HTTP Live Streaming (HLS)}
As the name suggests, HLS is an adaptive streaming protocol that delivers content over HTTP/HTTPS. The \textit{HLS Architecture} \cite{hls-arch} essentially involves three components-
\begin{enumerate}
    \item The Streaming Server
    \item The Distribution Component
    \item The Streaming Client
\end{enumerate}

\begin{figure}[H]
    \centering
    \includegraphics[width = 0.25\textwidth]{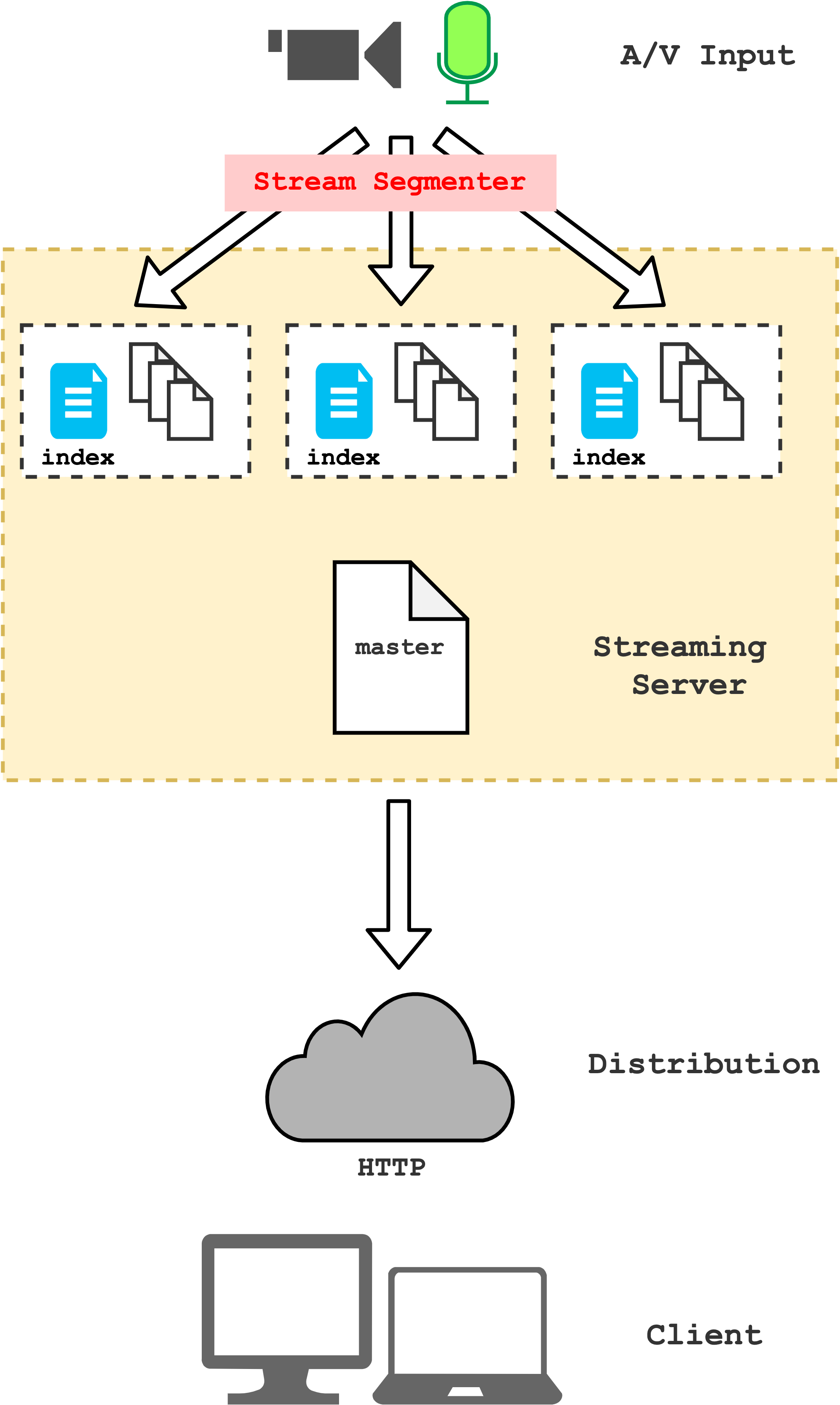}
    \caption{HLS Architecture}
    \label{hls-arch}
\end{figure}

A typical configuration (Figure \ref{hls-arch}) consists of a hardware encoder that encodes Audio/Video input into MP3/H.264 and encapsulates it into an MPEG-2 transport stream. A software segmenter then divides the stream into chunks(.ts files) of equal duration and creates an index file \texttt{index.m3u8} that contains links to those chunks. This process is carried out for each encoding of the A/V stream and a master index file, also called the \textit{manifest} is generated and usually named \texttt{master.m3u8}. The manifest identifies and points to the different index files available for that particular A/V stream. This manifest is then served by the streaming server over HTTP to the client which then selects the suitable encoding based on the resources available and requests the index file of that encoding. Once the index is received, the client sequentially makes requests for the chunks, enabling playback on its device. When a network change is detected, a lower bitrate encoded index file is requested in order to retrieve the next chunk, for continuous playback.

\subsection{Digital Rights Management (DRM)}
\label{drm}
DRM can be thought of as a digital lock or as protections in place to secure proper usage of proprietary technologies and copyrighted works\cite{DRMBook}. Although enforced in many countries through licensing agreements and laws such as the Digital Millenium Copyright Act (DMCA)\cite{DMCA} which criminalise circumvention, their efficacy and ulterior motives have been the subject of constant debate. A discussion on these technicalities is however not relevant to this paper, where we will instead choose to focus upon the DRM techiques used primarily by OTT service providers.
Most streaming services such as Netflix, Hulu, Amazon Prime require playback devices to support some form of DRM. Common choices fir DRM schemes are Microsoft PlayReady\cite{playready}, Apple FairPlay\cite{fairplay}, Adobe PrimeTime\cite{primetime}, Marlin\cite{marlin} and Google Widevine\cite{widevine}. Most of these DRM schemes  \textit{atleast} provide browser support through Content Decryption Modules\cite{EME} (CDMs) which follow the Encrypted Media Extensions (EME)
\cite{EME} specification which in turn is implemented by all major browsers today. This uniformity has resulted in an unprecedented ease of implementing basic content protection at an efficient cost. An example would be Google's Shaka Player\cite{shaka}: an open source player which can be integrated into a project with relative ease, which uses the Widevine DRM scheme and supports adaptive streaming over MPEG-DASH\cite{dash} and HLS.

\subsection*{Earlier Work on DRM}
Breaking DRM protection has been the focus of all kinds of hacker groups ever since the popularity of commercial software grew, giving rise to the so called pirate hubs which are still popular today. Right from spoofing KMS systems for acquiring Windows licenses to patching AAA titles deploying the Denuvo Anti Tamper\cite{denuvo} system, the community has been witness to some rather creative albeit illegal ways of stealing content over the years.
Acacdemic attention to the problem of breaking DRM systems \cite{Wyang,Kindle,ACMDRM} however, has proved to be rather mild. To the best of our knowledge, our work is the first such analysis of OTT Indian music streaming webapps. We did however take inspiration for the subject from the work done by Wang et. al \cite{Wyang} on automatically bypassing DRM systems and for a way to present our findings, we looked at Kumar et. al's \cite{UPI} work on analysing UPI apps in India.

\section{Spotify: Demonstrating Widevine}
\label{spotify}
 
Having been baffled by the results of our preliminary investigations into Wynk, we were curious to see if this trend was followed across the board by even the big players in the game and hence we decided to focus our attention on Spotify. We were quite satisfied to observe that Spotify proved resilient to the basic reversing techniques that had proved fatal for the other services in terms of content security.
However we must clarify, we do not claim that Spotify is infallible, just that it would require more effort than what was put in for all the others \textit{combined}. Here we present a high level overview, explaining how Spotify protects its content while streaming and also use this analysis later to highlight the deficiencies in the other protocols. 

\subsection{Methodology}
We decided to target the \textit{Spotify Web Player} as it was clearly suited for comparison with the other services. By monitoring network requests made by the web player and using a combination of static and dynamic analysis of the client-side Javascript modules, we were able to piece together the inner workings. The documented architecture\cite{widevine-arch} for Widevine was also heavily referred to in order to establish context.

\subsection{Summary Of Findings}
Spotify\footnote{For future reference, unless explicitly stated otherwise, a reference to Spotify refers to the Spotify Web Player} is currently using Widevine Level L3 to implement DRM for it's content which is streamed to a modified version of the Shaka player that uses the HTML5 Media Source Extensions to interact with the CDM.\footnote{For a detailed explanation, refer to the EME documentation\cite{EME}} 

The CDM is a \textit{precompiled} binary\footnote{An open source CDM or OCDM can be viewed in the Chromium Project's source, however the Chrome CDM is closed source \cite{DRMSupport}}, implemented as a shared library \textit{(}\texttt{libwidevinecdm.so} \textit{in Linux} for Google Chrome and as a plugin for Mozilla Firefox.

\begin{figure}
    \centering
    \includegraphics[width=0.5\textwidth]{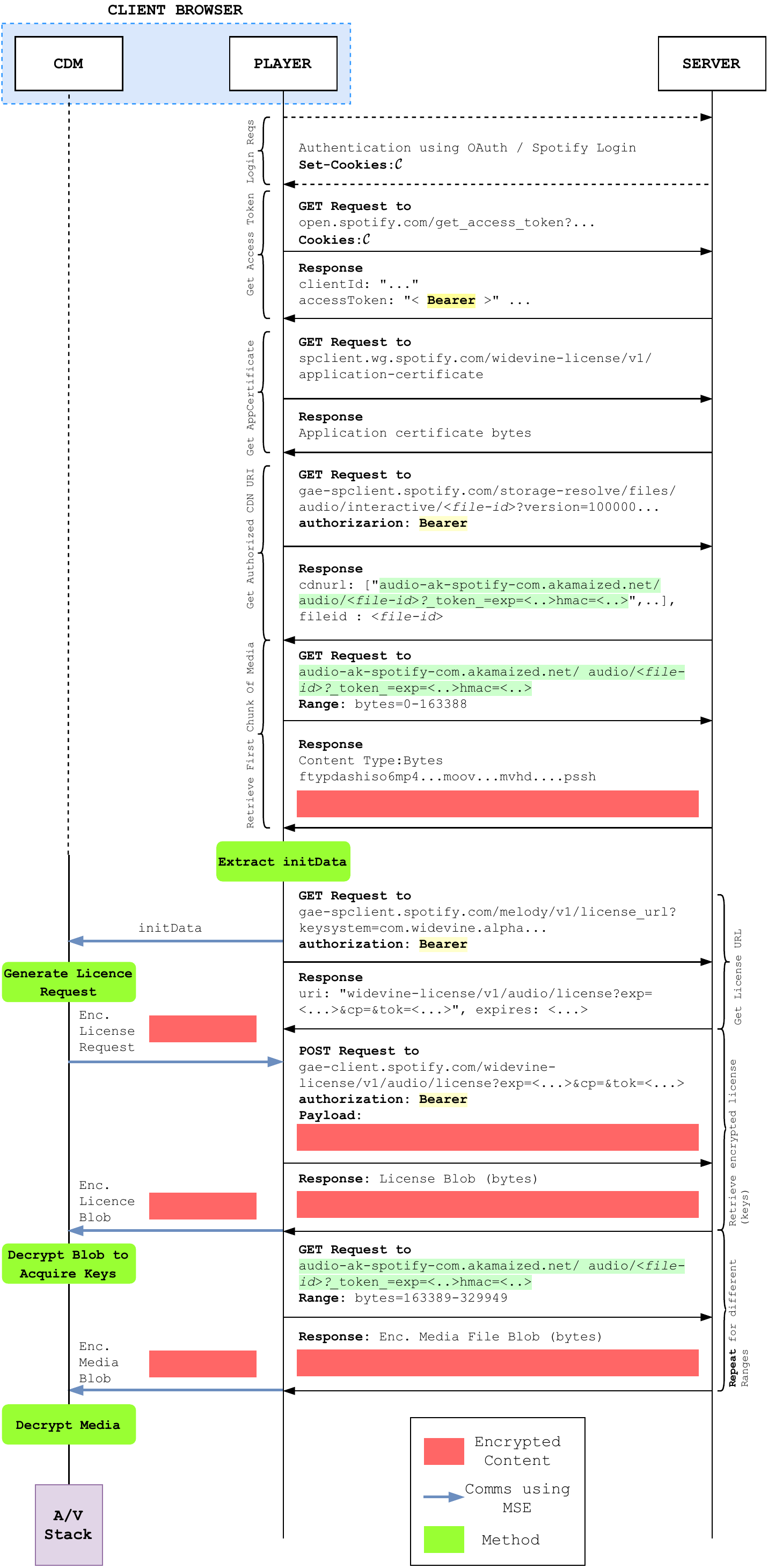}
    \caption{Spotify's Content Retrieval Protocol - Widevine L3 \textit{(Reconstructed)}}
    \label{fig-spotify}
\end{figure}

Coming to the retrieval, Figure (\ref{fig-spotify}) depicts the protocol followed. We would like to point out that we have intentionally left out the exact details in some parts of the protocol in the interests of keeping the description brief.

\begin{enumerate}
    \item \textbf{Login} This is the first part of the protocol that a client encounters when trying to start playback. Streaming does \textit{not} start until user identification and authentication is done. There are numerous options for authentication using OAuth but all of them effectively end up setting identification cookies on the client. Let us denote these cookies by $\mathcal{C}$. These cookies are later used to setup and maintain a player \textit{state} which is used to track playback, sync multiple devices, gather insights etc.
    
    \item  \textbf{Acquire Access Token} An access token is requested from the spotify servers using the cookies $\mathcal{C}$. As in the protocol, we shall refer to this token by \texttt{Bearer}. \texttt{Bearer} is an authenticated token that has a long expiration time and is required by and subsequently used for most operations of the Spotify client. 
    
    \item \textbf{Get Resource URI} To actually retrieve the media file from the Content Distribution Network (CDN), we require an authorized URI which permits access to the content on the server. This URI is obtained from the Spotify servers by making a request and leveraging the \texttt{Bearer} token as authorization. If the \texttt{Bearer} token is valid, the server responds with a list of multiple URIs \textit(We assume for redundancy).
    
    \item \textbf{Retrieving the First Chunk Of Data} According to the widevine specification, the first chunk of data is used to gather licensing information for subsequent decryption of content. Having obtained the CDN URI in the previous request, the player requests the first chunk of data of a certain size by setting the \texttt{Range} header in the request. The server response which contains a chunk of the media file (\textit{distinguished from its header}) is used to extract initial licensing information called \texttt{initData}. \texttt{initData} is then passed on to the Content Decryption Module (CDM).
    
    \item \textbf{Obtaining the License} The fragmented\footnote{See \texttt{moof} boxes \cite{moofbox} for information on format of the media fragments} media chunks retrieved from the servers are encrypted using \texttt{AES-128} in \texttt{CTR} mode. Hence, in order to initlialise playback, decryption of these media needs to be performed. The information (keys, initlialisation vectors) needed for decryption is included in the \textit{license}. Based on the \texttt{initData}, the CDM generates an \texttt{encrypted} license request and passes it to the player. The player then relays this request to a license URI that was obtained asynchronously. If the request and its payload is valid, the server responds with an \textit{encrypted} response that is relayed by the player to the CDM. The CDM decrypts the response to obtain the license.
    
    \item \textbf{Playback} Once the licensing information has been obtained, 10 second chunks are downloaded from the servers and passed to the CDM which decrypts those chunks and passes them to the Audio/Video Stack for playback.

\end{enumerate}

\subsection{Lessons Learnt}
Spotify does a few things very well for protecting their content as it is streamed to a client's device. For the sake of establishing standard practices, we highlight a few of them below -
\begin{enumerate}
    \item \textbf{Mandatory User Identification} The login process forces a client to identify itself in order to use the services. Spotify, while providing flexibility of login options with OAuth also implements reCAPTCHA protection against bots. In addition to this, Spotify can track each user's activity which could potentially be used to recognise malicious use.
    \item \textbf{Streamed Content is Encrypted} To prevent \emph{streamripping} \textit{(discussed in Section \ref{streamripping})}, content stored on the servers is encrypted.
    \item  \textbf{No Hardcoded Keys} The keys for decryption of the streamed content are not hardcoded in the files that the user has direct access to.
    \item \textbf{License Information is Invisible to the Player} The license information passed between the CDM and the server is encrypted and hence is not accessible to the user.
    \item \textbf{Content Decryption Module} The CDM is theoretically the weakest part of the protocol. However in terms of \textit{usable}/\textit{practical} security, since it is closed source binary, it offers a basic level of protection against direct observation of the decrypted content, but is theoretically vulnerable to black box cryptanalysis techniques and some implementation level exploits. L2 and L1 level Widevine attempt to mitigate this vulnerability by having the decryption occur in a \textit{Trusted Execution Environment(TEE)}\cite{DRMExplained}.
    
\end{enumerate}

Now that a benchmark has been established, we proceed to present an analysis of the four biggest OTT music streaming services in India, in the process highlighting security gaffes where DRM is concerned.
\section{Case Studies}
\label{case-study}
This section forms the basis of the work done in this paper. We present here, a reconstruction of the streaming protocols used by the four biggest \textit{(by subscriber base)} music streaming services in India, in view of formulating an exploit to steal their content. The reconstruction of these protocols involved reverse engineering the Javascript modules executing on the client browser using static and dynamic techniques such as code de-obfuscation, debugging etc., observing network packets using Burp\cite{ref_burp} and a fair amount of intuition. In all cases, we were able to \emph{completely replicate} the protocols in order to get access to the audio content using these standard techniques of reverse engineering. Some code obfuscation aside, none of these services used industry standard DRM and could be broken with minimal effort by a dedicated attacker.

Given below is a summarised analysis of Airtel Wynk, JioSaavn, Gaana and Hungama. Using this analysis we were able to write scripts to automatically steal content. In the interest of keeping the descriptions concise, we have deliberately sacrificed rigorous function definitions in favour of broad descriptions of what those functions do, while illustrating a protocol. The token/variable names mentioned are also similar to their names in the actual JS code. For a detailed description refer to Appendix~\ref{pseudocode}.  The implementation details are furnished in Section \ref{impl}.
\subsection{The Curious Case Of Wynk}
Airtel Wynk Music was the first service that we came across that had serious flaws in their content security mechanism. The flaws were such, that we were able to write scripts in order to automatically steal content at the highest available quality. The protocol diagram in Figure \ref{wynk1} describes the working of Wynk prior to our disclosure.
\begin{figure}
    \centering
    \includegraphics[width=0.45\textwidth, trim=0 30 0 0,clip]{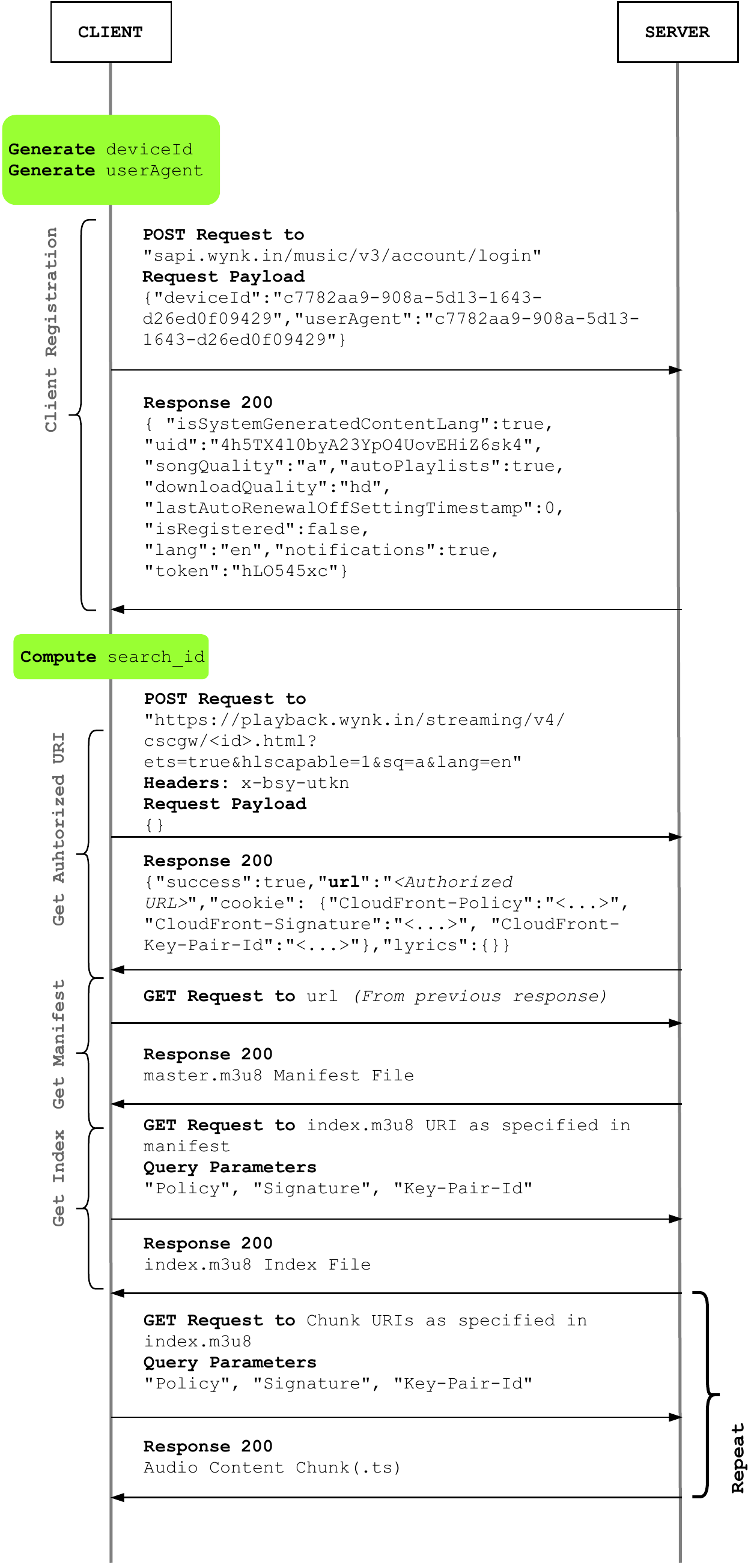}
    \caption{Content Retrieval Protocol - Wynk \textit{(Reconstructed)}}
    \label{wynk1}
\end{figure}
\subsubsection*{The Protocol in A Nutshell}
\begin{figure}
    \centering
    \includegraphics[width=0.45\textwidth]{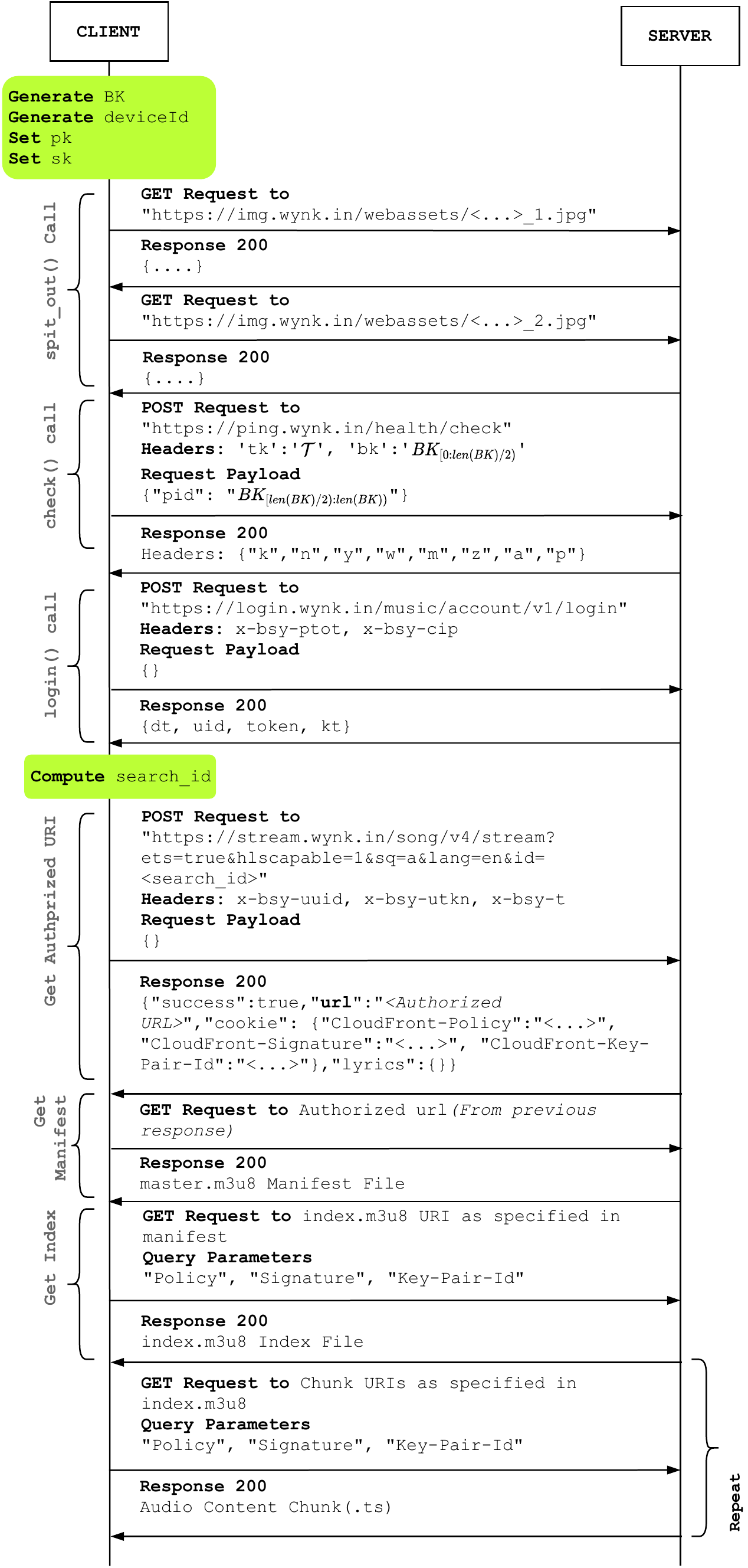}
    \caption{Revised Content Retrieval Protocol - Wynk \textit{(Reconstructed)}}
    \label{wynk2}
\end{figure}

\begin{itemize}
    \item \textbf{Client Registration} The client is identified to the server using a \texttt{POST} request containing the \texttt{deviceId} and \texttt{userAgent} parameters in the payload. These parameters are set by the client and appear to be random in nature. Our observation was that persisting the values for these parameters had no effect on the execution of the protocol. As a response to this request, the server replies with values for \texttt{uid} and \texttt{token}.
    \item \textbf{Compute Search Id For Resource} A \texttt{search\_id} was computed based on the song URL through a combination of string operations and table lookups.
    \item \textbf{Acquire Authenticated URI} A \texttt{POST} request is made to retrieve the authenticated URI for content retrieval from the CDN. 
    Using \texttt{token} as the key, a \texttt{SHA1-HMAC} of a string containing the \texttt{search\_id} is generated. The Base64 encoded value of this HMAC is assigned to a request header \texttt{x-bsy-utkn} after appending it to the \texttt{uid}. The result of this request is a URL with a set of signed cookies which we will refer to as \textit{CloudFront Cookies}.
    \item \textbf{Acquire Manifest} On making a request to the URI obtained as a response in previous request, the server responds with the manifest file which contains URIs to the various \texttt{index.m3u8} files available.
    \item \textbf{Acquire \texttt{index.m3u8}} Using the URI for the index file of highest quality available, a request is made with query parameters being set using the \textit{CloudFront Cookies} obtained previously. A successful response from the server gives us the \texttt{index.m3u8} file of our choice. 
    \item \textbf{Getting Content} By making \texttt{GET} requests to the chunk URIs present in the index file and setting the appropriate query parameters, the client starts recieving \texttt{.ts} media files from the server. By appending those files in order, the complete audio file is obtained.
\end{itemize}
Following our disclosure, Wynk made certain changes to their protocol that are listed below -
\begin{enumerate}
    \item A code obfuscation scheme was introduced that replaced function/variable names, identifiers with what were essentally array lookups. The array used for lookup was included in the source code which rendered the obfuscation useless.
    \item The client registration process was redesigned and a time window was introduced using Time Based OTPs (TOTPs)\cite{totp}.
\end{enumerate}
However, the the content retrieval part of the protocol remained the same. The introduced changes only served to complicate the process of getting the authenticated URIs for the CDN. Moreover, content was still being streamed without encryption\footnote{When we say \textit{without encryption}, we refer to the fact that after the decryption from the HTTPS layer and gzip unpacking has occurred, the audio content is \textbf{directly playable (unencrypted)}}. The revised protocol is described in Figure \ref{wynk2}. Needless to say, we did not face any difficulties in breaking Wynk once more.

\subsubsection*{The New Protocol in a Nutshell}
\begin{itemize}
    \item \textbf{Initialisation Of Identifiers} The first part of the protocol involved the generation of certain tokens, namely, \texttt{BK, deviceId, pk \& sk}. \texttt{pk} and \texttt{sk} were values that were hardcoded into the source code while \texttt{BK} and \texttt{deviceId} were generated using the \textit{epoch time} and a \textit{pseudorandom number}
    \item \textbf{\texttt{spit\_out(BK, deviceId)}} Two requests are made to the server using the outputs of this method which basically does some intermixing of the strings \texttt{deviceId} and \texttt{BK}. The ouptut strings are then appended with \texttt{"\_1.jpg"} and \texttt{"\_2.jpg"} and treated as endpoints for requests. Now, we are not entirely sure why the image extensions are used in particular, however we can confidently say that the response to those requests has no further use. That being said, if either of those requests are not made, the protocol fails subsequently,
    \item \textbf{\texttt{check() \& login()}} These functions are named after the endpoints to which requests are made. A successful response to the check endpoint returns several parameters which are used to compute the values of certain headers in the request to the login endpoint. A successful response from the \texttt{login} endpoint contains the parameters \texttt{dt, uid, token, kt} among others.
    \item \textbf{Compute Search Id} This method did not change compared to the previous deployment of Wynk
    \item \textbf{Acquire Authenticated URI} The values received in the previous step are used to set the headers for another request as follows - 
    \begin{itemize}
        \item \texttt{x-bsy-uuid} $\leftarrow$ \texttt{dt}
        \item \texttt{x-bsy-utkn} similar computations as before\footnote{The changes can be observed in the Appendix}
        \item \texttt{x-bsy-t} $\leftarrow$ \texttt{AES(kt, TOTP(dt||sk, 600, 6))}\footnote{\texttt{6 digit TOTP generated with a window of 600 seconds.} CryptoJS implementation of AES used}
    \end{itemize}
    This \texttt{POST} request if successful returns the \textit{CloudFront Cookies} with a URI. The rest of the protocol follows identically to the previous version of Wynk.
\end{itemize}

It is pretty evident from the analysis that Wynk went to greater lengths to complicate the retrieval mechanism post disclosure, however they failed to address the crux of the problem.  
\subsection{JioSaavn Joins The Jam}
With the findings from our work on Wynk, we were inspired to look into other platforms to test if the situation found with Wynk was a general norm among established players.
JioSaavn is the 2\textsuperscript{nd} most popular India based music streaming service in terms of number of subscribers.
It took some vigilant effort to get to the media content but once the relevant execution path was found, piecing together the protocol was found to be extremely easy and straightforward.

\begin{figure}[htbp]
    \centering
    \includegraphics[width=0.45\textwidth]{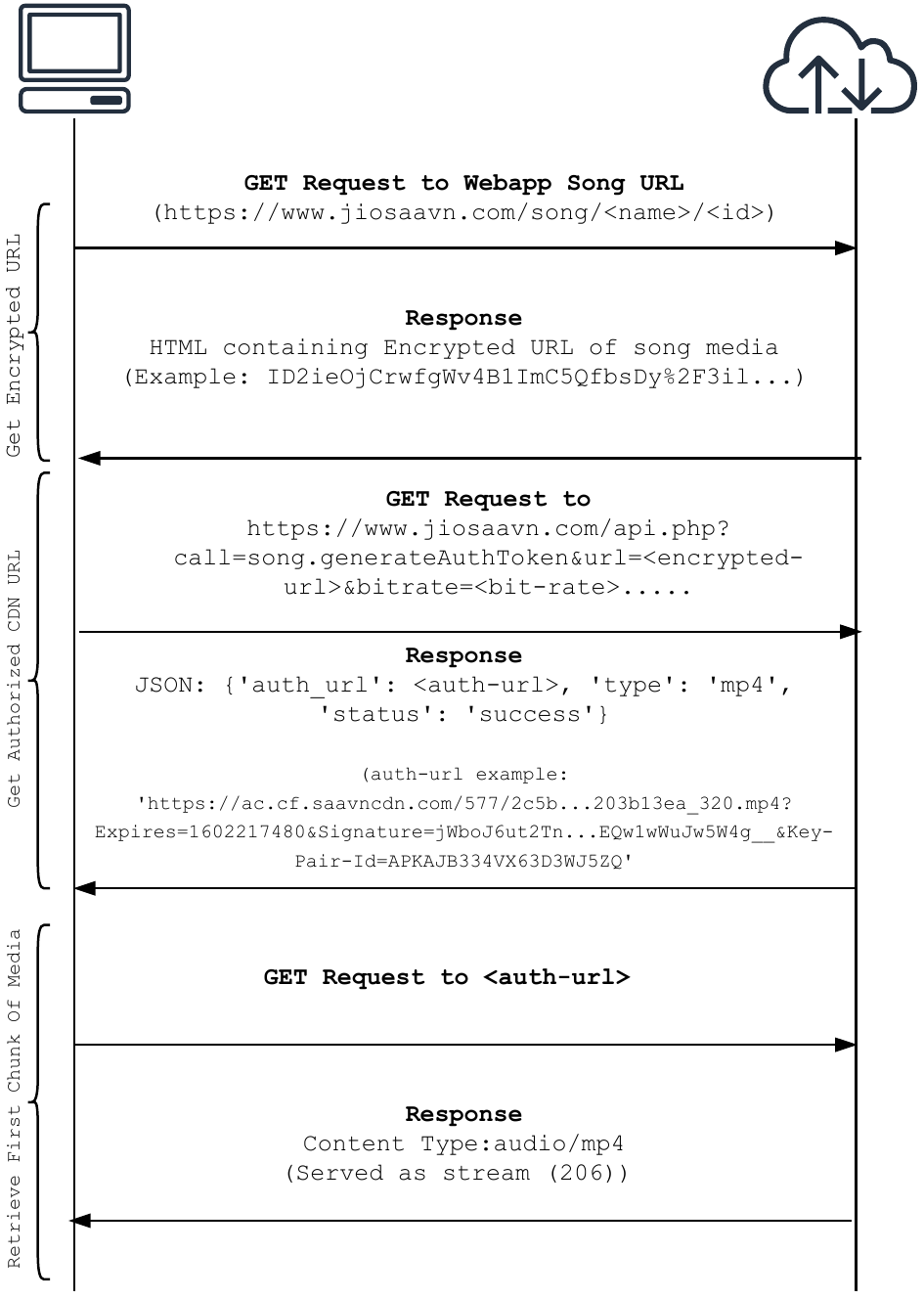}
    \caption{Content Retrieval Protocol - JioSaavn \textit{(Reconstructed)}}
    \label{saa1}
\end{figure}  
\subsubsection*{The Protocol in A Nutshell}
\begin{itemize}
    \item \textbf{Acquiring Song Info} Interestingly, the metadata related to the song is served within an HTML response.
     It is found within the \emph{JavaScript} variable,\\
     \texttt{window\.\_\_INITIAL\_DATA\_\_}.\\
     Parameters that are essential for fetching the media content are \texttt{encrypted\_medial\_url} and \texttt{perma\_url}.
    \item \textbf{Generate Auth Token} A \texttt{GET} request is made to
     \url{https://www.jiosaavn.com/api.php?call=song.generateAuthToken&url=<encrypted_media_url>&bit_rate=<bit_rate>}
     to obtain the authorised URL that is used to fetch the media from the CDNs. The relevant parameters are
      \texttt{url} which is the \texttt{encrypted\_media\_url} discussed above and \texttt{bit\_rate} which takes the
       values \texttt{"128", "320", "64", "32", "16"}. The response contains \texttt{auth\_url} which is verified by the CDN to authorize a request.
    \item \textbf{Downloading Media} A \texttt{GET} request is made to \texttt{auth\_url} to finally retrieve the relevant media.
     This URL is sufficient for authorization.
\end{itemize}

\subsection{Getting Gaana}
\begin{figure}
    \centering
    \includegraphics[width=0.4\textwidth,trim=0 30 0 0,clip]{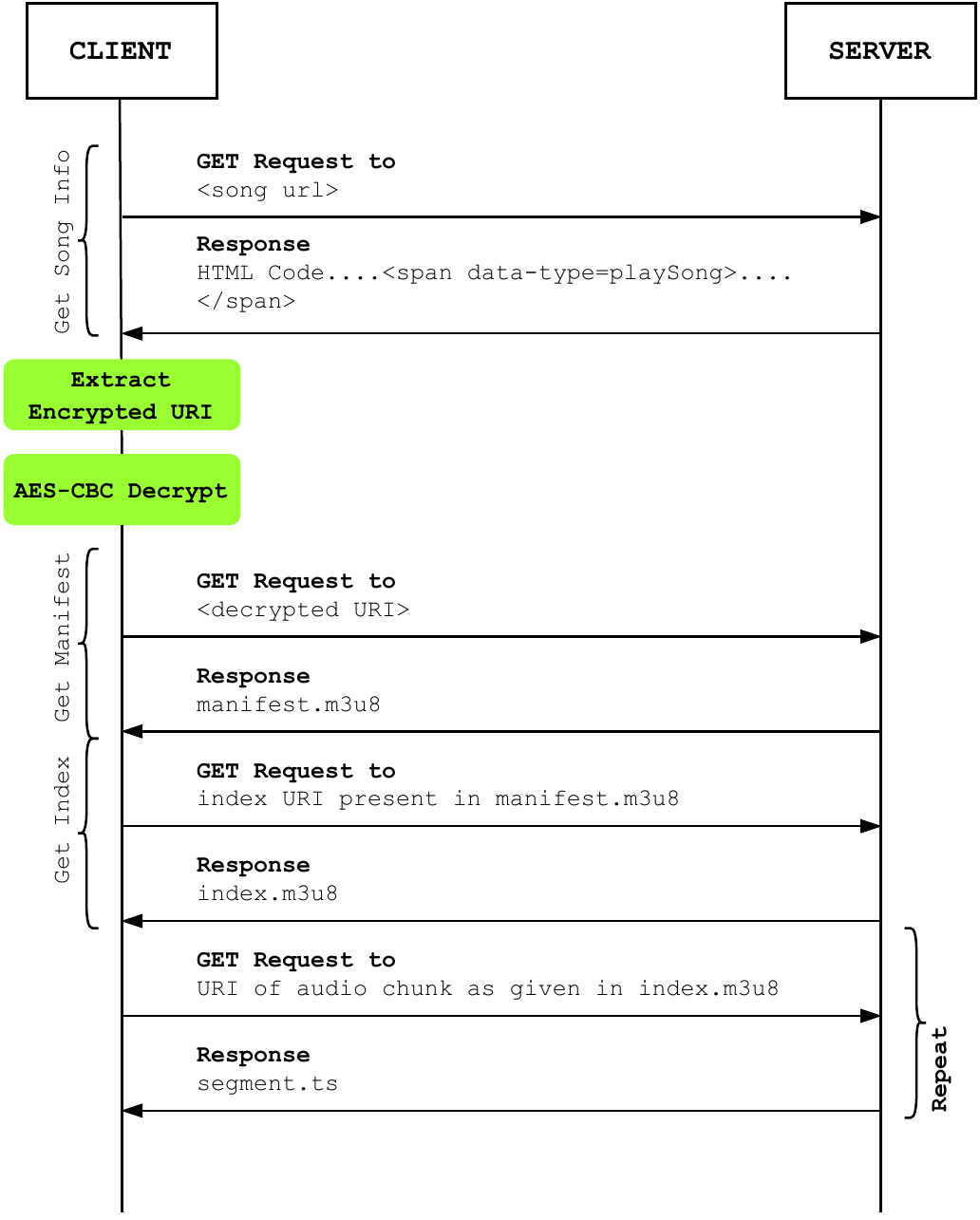}
    \caption{Content Retrieval Protocol - Gaana (\textit{Reconstructed})}
    \label{gaana}
\end{figure}

Gaana was the third service that we looked at. Figuring out the protocol was easy as Gaana had no code obfuscation and at least for the non-logged in user, did not rely on cookies at all. Figure \ref{gaana} demonstrates the working of the protocol.
\begin{itemize}
    \item \textbf{Getting Song Details} Where Wynk relied on interaction with the server to obtain an authorised resource URI, Gaana instead embeds all information as text in the \texttt{.html} code of the song's page. The current song information is a JSON string present in a \texttt{<span>} tag with \texttt{'data-type':'playSong'}. The \texttt{path} key contains \texttt{AES-CBC} encrypted URIs with \texttt{PKCS\#7} padding\cite{pkcs7} for various bitrates indexed as \texttt{high, medium, low}. 
    \item \textbf{Decrypting \texttt{path}} The \textit{key} and \text{initialization vector} is \emph{hardcoded} in the JS files and we use those values to decrypt and obtain the authorized URI.
    \item \textbf{Acquire Manifest} A request to the authorised URI returns the \texttt{manifest.m3u8} which contains the URI for \texttt{index.m3u8}
    \item \textbf{Playback} The \texttt{index.m3u8} file contains URIs for all chunks of the audio. After iterating throught this file and making requests for all chunks (\texttt{.ts} files), we can append them together to obtain the complete audio.  
\end{itemize}
\subsection{Hunting Hungama}
Hungama is yet another popular music streaming service in India. We explored its content serving mechanism and found it to be pretty similar to JioSaavn and reverse engineered the following protocol.
\begin{figure}
    \centering
    \includegraphics[width=0.45\textwidth,trim=0 30 0 0,clip]{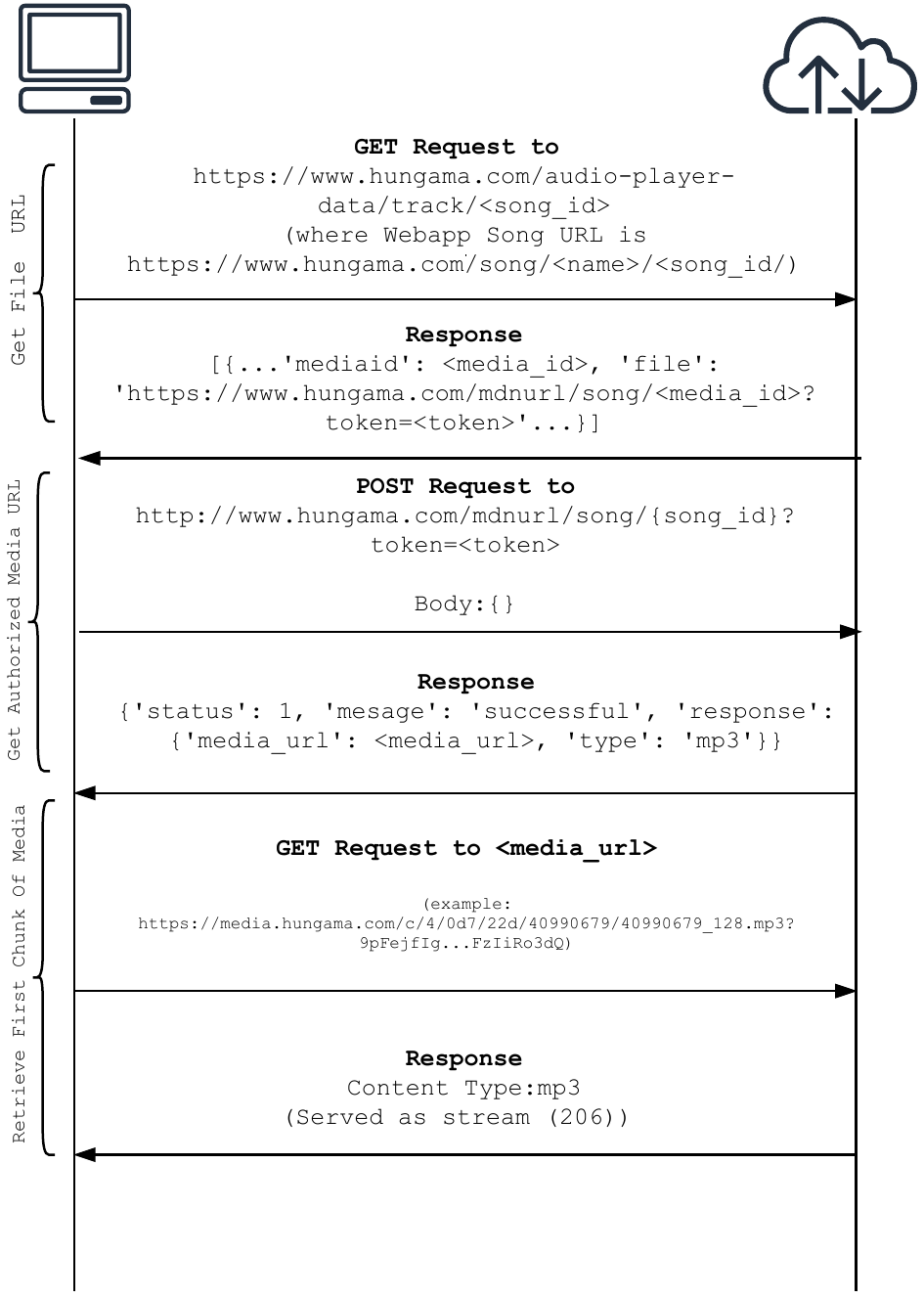}
    %\Description[short description]{long description}
    \caption{Content Retrieval Protocol - Hungama \textit{(Reconstructed)}}
    \label{hungama}
\end{figure}  
\subsubsection*{The Protocol in A Nutshell}
\begin{itemize}
    \item \textbf{Audio Player Data} A \texttt{GET} request is made to \url{https://www.hungama.com/audio-player- data/track/<song_id>} to fetch the metadata of the song. The \texttt{song\_id} parameter required for this request is obtained from the WebApp URL of the song which is of the form \url{https://www.hungama.com/song/<name>/<song_id/}.
    The response contains the relevant values \texttt{media\_id} and \texttt{file}. The \texttt{file} URL contains the parameter \texttt{token}
    \item \textbf{Fetching Media Url} A \texttt{POST} request with an empty body is made to \url{http://www.hungama.com/mdnurl/song/{song_id}?token=<token>}.
    The response contains \texttt{media\_url} which is the final URL that is used to retrieve the media. The bit rate can be chosen by setting the \texttt{hcom\_audio\_qty} parameter to one of \texttt{"high", "low", "medium"} in the \texttt{Cookie} header.
    \item \textbf{Downloading Media} A \texttt{GET} request is made to \texttt{media\_url} to download the relevant media. This URL is sufficient for authorization.
\end{itemize}

\section{Discussion}
\label{discussions}

\subsection{Comparative Analysis}
Among the case studies showcased in this work, reversing the protocol for Wynk was the most challenging in terms of effort due to the intricacies and complexity of the implementation.
Yet, in the end it turned out to be a matter of getting through the different layers of obfuscation which didn't really have any theoretical security guarantees. When compared to Wynk, the other 3 services, JioSaavn, Gaana and Hungama had simpler mechanisms for content serving which were reversed quite effortlessly. The patch implemented by Wynk after our disclosure shows that they are indeed concerned about protecting their content and so it would be quite interesting to
see how these platforms embrace DRM in the future instead of working on stop-gap solutions which only delay the inevitable. \\
We present a summary of the best/worst practices in Table \ref{table-comparison} based on our investigations.

\begin{table}[htbp]
	\resizebox{\linewidth}{!}{%
		\renewcommand{\arraystretch}{1}
		\begin{tabular}{lccccc}
			\hline
			\multicolumn{1}{c}{\textbf{Practice}} &
			\textbf{Spotify} &
			\textbf{Wynk} &
			\textbf{JioSaavn} &
			\textbf{Gaana} &
			\textbf{Hungama} \\ \hline
			\begin{tabular}[c]{@{}l@{}}Mandatory User\\ Identification\end{tabular}        & \checkmark &            &  &            &  \\ \hline
			\begin{tabular}[c]{@{}l@{}}Streamed Content\\ Encryption\end{tabular}          & \checkmark &            &  &            &  \\ \hline
			Hardcoded Keys                                                                 &            & \checkmark &  & \checkmark &  \\ \hline
			DRM Scheme                                                                     & \checkmark &            &  &            &  \\ \hline
			\begin{tabular}[c]{@{}l@{}}Cookie Based \\ Authentication \\ Timeout\end{tabular} &
			\checkmark &
			\checkmark &
			&
			&
			\\ \hline
			\begin{tabular}[c]{@{}l@{}}Premium Content \\ Access Restrictions\end{tabular} & \checkmark &            &  &            &  \\ \hline
			\begin{tabular}[c]{@{}l@{}}Obfuscation/\\ Minification\end{tabular} &
			\checkmark &
			\checkmark &
			\checkmark &
			\checkmark &
			\checkmark \\ \hline
		\end{tabular}%
	}
	\caption{A Comparative Analysis of Practices}
\label{table-comparison}
\end{table}
\subsection{Why No DRM ?}
\label{drm-discussion}
What might be the possible reasons that could have led to such an oversight, we wonder. DRM protection is not exactly a new or novel concept and has been a part of the industry for a fair amount of time. This would imply that deploying these services with such primitive security measures was a conscious decision at its worst and a rushed one at its best.
A possible reason that comes to mind is the shaky compatibility of the discussed DRM schemes with current playback devices which could result in the alienation of a large chunk of the subscriber base. In a massive, competitive market such as India, this could potentially spell the demise of a competing streaming service at the hands of its rivals.
\subsection{An Inherent Problem of the Analog Hole~\cite{analoghole}}
\label{analog-hole}

For a considerable amount of time now, there has been a raging debate on the effectiveness of DRM solutions. The aptly named \textit{analog hole} is a problem inherent to the task of protecting Audio/Video Content and is something that is touted by DRM critics all the time and for good reason. Put simply, the analog problem can be illustrated as follows -
Imagine an ideal system that is completely secure in terms of communication and implementation. The client receives the encrypted content from the server and the decryption process occurs securely following which, the content is displayed by the screen or played by the speakers. The analog hole problem states that human perceivable analog signals can always be recaptured and re-encoded to a digital format, thereby nullifying all previous protection. As an example, a pirate can always re-capture video content by directing the content to a video card instead of a screen and similarly tapes/mics can be used to re-capture audio content.

The proponents of DRM however argue that since all analog to digital conversions are \textit{lossy}, one can never actually retrieve the original content by exploiting the analog hole. We leave it to the readers to decide which side of the fence they lean on.
\subsection{Attacks On Widevine}
Widevine is a DRM system developed by Google in order to protect content from misuse by the client. L3 level Widevine is supported by the latest versions of most major browsers and in our case, it is used by our benchmark - Spotify. When we chose to treat Spotify as the \textit{ideal} service with content protection we did so by blindly trusting Widevine based on its popularity. We attempt to rectify our assumptions in this section.

There have been multiple reports of researchers breaking L3 level Widevine. The first such claim was made by a security researcher \textit{David Buchanan} in a tweet~\cite{dbuchanan} in January 2019. This was followed by a blog from \textit{Fidus Information Security}~\cite{fidusinfosec} who claimed to have decrypted an episode of \textit{Stranger Things} from Netflix which used L3 Widevine. This group claims to have used a modified Differential Fault Analysis (DFA) approach to recover the keys used for decryption by the Widevine Module. Despite making claims, neither of the security researchers published working PoCs or exploits in order to prevent misuse. Speaking of misuse, a blackhat group called \textit{The Scene} also claims to have a working break~\cite{scene} which they use to pirate content off of Amazon Prime and Netflix.
\subsection{StreamRipping}
\label{streamripping}
In this section we discuss an alternate attack strategy that continues to be very popular among pirates~\cite{streamripinc}. This attack strategy, called \emph{StreamRipping} exploits the \textit{un-encrypted} nature of the streamed content within a browser. Essentially, once the decryption from the HTTPS layer has occured, if the content itself is not encrypted, it is visible to the browser and hence the attacker. This content can then be simply dumped to a file for later distribution - thus \textit{ripping from the stream}.

There are various tools and services which make it easy to \emph{StreamRip}.
An example would be the browser add-on, \emph{Audio Downloader Prime}~\cite{audprime} for \emph{Mozilla Firefox}. We found this tool to be quite effective in seamlessly capturing the content streamed from JioSaavn and Hungama where the media was being served as in response to a single request. In the case of Wynk, the chunks of the stream were detected and were retrievable as \texttt{.mp3} files. 

\subsection{Mitigation Strategies}
Following our analysis, we found most of the streaming services were guilty of malpractices (Refer Table \ref{table-comparison}). Further, our success in completely reverse engineering the protocols shows without a doubt that shallow patches will not prove to be secure in the long term. Short term strategies might include stronger obfuscation, reliance on precompiled binaries etc. but none of these techniques would stand the test of time. As of today, the best possible protection available is through DRM schemes like Widevine, Playready, Fairplay and given the ease of setting up these schemes, it might turn out to be the best possible long term strategy as well.

\section{Implementation Details}
\label{impl}

After much deliberation and thought over the fact that a working PoC could potentially be misused, we decided to exclude it from this work. As our experience with Wynk taught us, unless radical changes are made to those services, their protocols can potentially always be broken and data stolen on a massive scale. Our work illustrates this very fact. Ultimately we hope that these services deploy proper DRM measures and not a workaround patch that will perpetuate this game of cat-and-mouse. 

\section{Conclusion}
\label{conclusions}
In this paper we surveyed various OTT Music streaming services, particularly those of Indian origin and analysed their attack surface. Owing to their growing popularity as the de facto standard of media consumption, our observations led us to conclude that most streaming services were highly vulnerable to basic stream ripping and reverse engineering attacks. To present our case, we analyzed four most popular music streaming apps in India namely Wynk, Gaana, JioSaavn and Hungama and were successfully able to steal protected content without restrictions. In case of Wynk, we would like to emphasize the fact that even after the attacks had been disclosed, the patched version of their protocol was broken again using the same techniques, clearly showcasing the pitfalls of doing patchwork instead of adopting a systematic solution.  A detailed comparative study is furnished to show the extent of deviation from the state-of-the-art and possible mitigation strategies are also proposed. Through our work, we hope that such platforms take cognizance of the lax security measures in place and improve upon them.

\appendix
\section{Wynk Music Function PseudoCodes}
\label{pseudocode}
\subsection{Wynk v1}

\begin{algorithm}
    \KwIn{$url$\tcc*{Resource Url}}
    \KwOut{[$chunks$]\tcc*{Audio Chunk List}}
    \Begin(Initialisation){
        $deviceId$, $userAgent$ $\leftarrow$ Arbitrary strings \\
    }
    \Begin(Authentication){
        \tcc{Authorisation with the Wynk Servers to enable authenticated retrieval requests}
        $uid, token$ $\leftarrow$ register($deviceId, userAgent$)\\
        search\_id $\leftarrow$ get\_search\_id($url$)\\
        $C$ $\leftarrow$ request\_manifest($token$, search\_id)
    }
    \Begin(Retrieval){[$chunks$] $\leftarrow$ get\_song($C$)} 
    \caption{Client Side in Wynk 1.0}
    \label{wynk1}
\end{algorithm}

\begin{algorithm}
    \KwIn{$deviceId, userAgent$}
    \KwOut{$uid, token$}
    \KwData{Request Headers: $\mathcal{H}$}
    \Begin{
        $url$:= \url{"https://sapi.wynk.in/music/v3/account/login"}\\
        payload := \texttt{\{"deviceId": "$deviceId$", "userAgent": "$userAgent$"\}}\\
        $uid, token$ $\leftarrow$ POST($url$, $\mathcal{H}$, payload)\\
        \Return $uid, token$
    }
    \caption{register() Function}
    \label{reg}
\end{algorithm}

\begin{algorithm}
	\KwIn{$url$}
	\KwOut{$search\_id$}
	\KwData{\texttt{STATIC\_CPMAPPING[ ]}}
	\Begin{
		\tcc{Each URI on Wynk has the following format -
			\texttt{\textbf{"producerid\_string"}}
		}
		Extract \texttt{$producerid$\_$string$} from $url$\\
		$id$ $\leftarrow$ \texttt{STATIC\_CPMAPPING[$producerid$}] \\
		\Return ($id$ $||$ \texttt{"\_"} $||$ \texttt{$string$})		
	}
	\caption{search\_id() Function}
	\label{searchid}
\end{algorithm}

\begin{algorithm}
    \KwIn{$search\_id, token, uid$}
    \KwOut{Signed CloudFront Parameters}
    \KwData{Request Headers: $\mathcal{H'}$}
    \Begin{
        \tcc{Generating header x-bsy-utkn}
        prefix := \url{"/streaming/v4/cscgw/"}\\
        suffix := \url{".html?ets=true&hlscapable=1&sq=a&lang=en{}"}\\
        subdomain := \url{"https://playback.wynk.in"}\\
        $msg$ := $\texttt{"POST"}||$ prefix $||$ $search\_id$ $||$ suffix\\
        $digest$ $\leftarrow$ SHA1\_HMAC($token$, $msg$)\\
        $\mathcal{H'}$[x-bsy-utkn] $\leftarrow$ $uid||\texttt{":"}||\texttt{Base64Enc(}digest\texttt{)}$\\
        $auth\_url$ $\leftarrow$ subdomain $||$ prefix $||$ $search\_id$ $||$ suffix\\
        $C$ $\leftarrow$ POST( $auth\_url$, $\mathcal{H'}$, payload: $"\{\}"$ )\\
        \Return $C$
    }
    \caption{request\_manifest() Function}
    \label{reqman}
\end{algorithm}
% \clearpage
\begin{algorithm}
    \KwIn{$C$ \tcc*[]{Response Object Containing Authenticated URIs \& Signatures}}
    \KwOut{[$chunks$]}
    \Begin{
        manifest\_url $\leftarrow$ Extract manifest file URI from $C$\\
        manifest\_file $\leftarrow$ GET(manifest\_url)\\
        index\_uri $\leftarrow$ Identify and extract highest quality \texttt{index.m3u8} URI\\
        index\_file $\leftarrow$ GET(index\_uri)\\
        chunks = [ ]\\
        \ForAll{chunk\_url in index\_file}{
            $tmp$ $\leftarrow$ GET($chunk\_url$)\\
            chunks.push[$tmp$]
        }
        \Return chunks
    }
    \caption{get\_song() Function}
    \label{alg5}
\end{algorithm}

\clearpage
\subsection{Wynk v2}

\begin{algorithm}
    \KwIn{$url$\tcc*{Resource Url}}
    \KwOut{[$chunks$]\tcc*{Audio Chunk List}}
    \Begin(Initialisation){
    $BK$ $\leftarrow$ gen\_bk($\mathcal{T}$,$\mathcal{R}$) \\
    $deviceId$ $\leftarrow$ gen\_random\_id($\mathcal{R}$)\\
    $pk$ = \texttt{Base64enc(https://sapi.wynk.in/music})\\
   
    $sk$ = \texttt{51ymYn1MS}
    }
    \Begin(Authentication){
        \tcc{Authorisation with the Wynk Servers to enable authenticated retrieval requests}
    spit\_out($BK$, $deviceId$)\\
    $U:=\{k,n,y,w,m,z,a,p\}$ $\leftarrow$ check($BK$, $\mathcal{T}$)\\
    $M:=\{dt,uid, token, kt, ...\}$ $\leftarrow$ login($U$, $\mathcal{T}$)\\
    $C$ $\leftarrow$ request\_manifest($url$, $token$, $dt$, $sk$)\\
    }
    \Begin(Retrieval){[$chunks$] $\leftarrow$ get\_song($C$)} 
    \caption{Client Side in Wynk 2.0}
    \label{algwynk2}
\end{algorithm}

\begin{algorithm}
    \KwIn{$BK$, $deviceId$}
    \Begin{
        $d_1, d_2$ $\leftarrow$ $deviceId_{[0...36)}, deviceId_{[36...72)}$\\
        In $d_1, d_2$ replace \text{"$-$" $\rightarrow$ ""}\\
        $d_3, d_4$ $\leftarrow$ mix\_it($d_1$, $BK$), mix\_it($d_2$, $BK$)\\ 
        $url$ := \url{"https://img.wynk.in/webassets/"} \\
        GET($url || d_3 || \texttt{"\_1.jpg"}$)\\
        GET($url || d_4 || \texttt{"\_2.jpg"}$)\\
    }
    \caption{spit\_out() Function}
    \label{spitout}
\end{algorithm}

%\subsubsection{\texttt{mix\_it() Function Definition}}
%\begin{verbatim}
%    def mix_it(a, BK):
%        b = list(a)
%        t = list(BK)
%        e = []
%        x = 0
%        n = 0
%        for i in range(31):
%                if i%2 == 0:
%                        e.append(b[x])
%                        x += 1
%                else:
%                        e.append(t[n])
%                        n += 1
%        e.append(''.join(t[n:len(t)-n-1]))
%        n = len(t)-n-1
%        for i in range(33,65):
%                if i%2 ==0:
%                        e.append(b[x])
%                        x+=1
%                else:
%                        e.append(t[n])
%                        n+=1
%        return ''.join(e)
%    \end{verbatim}

\begin{algorithm}
    \KwIn{$BK$}
    \KwOut{k,n,y,w,m,z,a,p}
    \KwData{Request Header: $\mathcal{H}$}
    \Begin{
        $url$:= \url{"https://ping.wynk.in/health/check"}\\
        $\mathcal{H}$[tk] $\leftarrow$ $\mathcal{T}$\\
        $\mathcal{H}$[bk] $\leftarrow$ $BK_{[0, len(BK)/2)}$\\
        $p$ $\leftarrow$ $BK_{[len(BK)/2), len(BK))}$\\
        $U$ $\leftarrow$ POST($url$, $\mathcal{H'}$, payload:$\{``pid":``p"\}$)\\
        return $U$  
    }
    \caption{check() Function}
    \label{check}
\end{algorithm}

\begin{algorithm}
	\KwIn{${k,n,y,w,m,z,a,p}$}
	\KwOut{$dt,uid,token,kt,...$}
	\KwData{Request Header: $\mathcal{H'}$}
	\Begin{
		$url$:= \url{"https://login.wynk.in/music/account/v1/login"} \\
		$BS$ := $k||n||y||w||m||z||a||p$\\
		$\mathcal{H'}$[x-bsy-ptot] $\leftarrow$ $\mathcal{T}$\\
		\tcc{Generate \texttt{x-bsy-cip} from BS value}
		a $\leftarrow$ [ ], b$\leftarrow$ 0, t$\leftarrow$ 0\\
		\For(){$t \leq len(BS)-1$}{
			$e = 10(BS[t]) + BS[t+1]$\\
			\eIf{$e \leq 55$}{
				\eIf{$2 \not | \; b$}{
					a.push($200+e$)
				}
				{a.push($100+e$)}
				b++
			}{
				a.push($100+e$)
			}
		}
		$\mathcal{H'}$[x-bsy-cip] $\leftarrow$ concat($a$)\\
		$C$ $\leftarrow$ POST($url$, $\mathcal{H'}$, payload: $\{\}$)\\
		return $C$  
	}
	\caption{login() Function}
	\label{login}
\end{algorithm}

\begin{algorithm}
    \KwIn{$\{ url, dt, uid, token, kt\}$}
    \KwOut{Authenticated CloudFront Resource Parameters}
    \KwData{Request Header: $\mathcal{H''}$}
    \Begin{
        $\mathcal{H''}$[x-bsy-uuid] $\leftarrow$ $dt$\\
        \tcc{Generating header x-bsy-utkn}
        suffix := \url{"/song/v4/stream?ets=true\&hlscapable=1\&sq=a\&lang=en\&id="}\\
        search\_id $\leftarrow$ get\_search\_id($url$)\\
        $msg$ := $\texttt{"POST"}||$ suffix $||$ search\_id $||\texttt{"\{\}"}$\\
        $digest$ $\leftarrow$ SHA1\_HMAC($token$, $msg$)\\
        $\mathcal{H''}$[x-bsy-utkn] $\leftarrow$ $uid||\texttt{":"}||\texttt{Base64Enc(}digest\texttt{)}$\\
        \tcc{Generating x-bsy-t using Time Based OTPs and CryptoJS AES}
        $\mathcal{H''}$[x-bsy-t] $\leftarrow$ AES($kt$, TOTP($dt||sk$, 600, 6))\\
        \tcc{Send POST Request To Server}
        $X$ $\leftarrow$ POST($url$, $\mathcal{H''}$, payload: $\{\}$)\\
        return $X$

    }
     \label{requestmanifest}
     \caption{request\_manifest() Function}
\end{algorithm}
\clearpage
\bibliographystyle{ACM-Reference-Format}
\bibliography{bibfile}

\end{document}